\documentclass{article}
\usepackage{color}
\usepackage{amsmath}
\usepackage{amssymb}

\newcommand{\R}{\mathbb R}

\newcommand{\NN}{\mathbb N}
\newcommand{\bbeta}{\mbox{\boldmath $\beta$}}
\def\DH{\rm I\kern-1.5pt\rm H\kern-1.5pt\rm I}

\def\DR{\rm I\kern-1.45pt\rm R}
\def\DC{\kern2pt {\hbox{\sqi I}}\kern-4.2pt\rm C}

\newcommand{\cH}{{\cal H}}

\newcommand{\ba}{\begin{array}}
\newcommand{\ea}{\end{array}}
\newcommand{\be}{\begin{equation}}
\newcommand{\ee}{\end{equation}}
\newcommand{\bea}{\begin{eqnarray}}
\newcommand{\eea}{\end{eqnarray}}
\newcommand{\bi}{\begin{itemize}}
\newcommand{\ei}{\end{itemize}}
\newcommand{\sfrac}[2]{\mbox{$\frac{#1}{#2}$}}

\def\theequation{\arabic{section}.\arabic{equation}}

\begin{document}
\thispagestyle{empty}
\begin{flushright}
ITP--UH--07/11\\
\end{flushright}
\medskip

\begin{center}
{\bf \Large Integrable generalizations of oscillator and Coulomb systems \\[4pt]
via action-angle variables }\\
\vspace{0.5 cm} {\large
T.~Hakobyan$\;^{a,b}$,
O.~Lechtenfeld$\;^{c,}$\footnote{
    corresponding author; phone:+49 511 762 3667; fax:+49 511 762 3023;
    email: olaf.lechtenfeld@itp.uni-hannover.de},
A.~Nersessian$\;^{a}$,
A.~Saghatelian$\;^{a}$ and
V.~Yeghikyan$\;^{a,d}$}
\end{center}

$\;^a${\sl Yerevan State University,
1 Alex Manoogian St., 0025 Yerevan, Armenia}

$\;^b${\sl Yerevan Physics Institute,
2 Alikhanyan Br., 0036 Yerevan, Armenia}

$\;^c${\sl Leibniz Universit\"at Hannover,
Appelstr. 2, 30167 Hannover, Germany}

$\;^d${\sl INFN - Laboratori Nazionali di Frascati,
Via E. Fermi 40, 00044 Frascati, Italy}

\bigskip

\begin{abstract}
\noindent
Oscillator and Coulomb systems on $N$-dimensional spaces of constant curvature
can be generalized by replacing their angular degrees of freedom with a compact
integrable $(N{-}1)$-dimensional system.
We present the action-angle formulation of such models in terms of the
radial degree of freedom and the action-angle variables of the angular subsystem.
As an example, we construct the spherical and pseudospherical generalization
of the two-dimensional superintegrable models introduced by Tremblay, Turbiner
and Winternitz  and by Post and Winternitz.
We demonstrate the superintegrability of these systems and give their
hidden constant of motion.
\end{abstract}

\section{Introduction}
Prominent in the theory of integrable systems is the Liouville
theorem~\cite{arnold}, which states that any
$2N$-dimensional Hamiltonian system with $N$ mutually commuting
constants of motion is integrable. Besides, the theorem states that,
if the level surface of these constants of motion are
{\sl compact and connected manifolds\/}, then they are diffeomorphic
to $N$-dimensional tori. This enables one to introduce action-angle variables
$({\bf I},{\bf \Phi})$, so that the Hamiltonian depends
only on the action variables~${\bf I}$, which are constants of motion.
The formulation of an integrable system in terms of these variables
yields a comprehensive geometric description of its dynamics and is
a useful tool for developing perturbation theory~\cite{arnold,goldstein}.
Furthermore, action-angle variables indicate the (non)equivalence of different
integrable systems, since they admit only two kinds of freedom:

-- in the functional dependence of the Hamiltonian on the action variables,
  $H=H({\bf I})$;

-- in the domain of the action variables,
${\bf I}\in [{\bbeta}^{-}, {\bbeta}^{+}]$.\\
Besides the standard textbook problems such as the harmonic oscillator
or the Kepler potential, action-angle variables seem to be absent in the
literature for the vast variety of known integrable models, such as
 integrable
systems in a curved geometry~\cite{sch,higgs} or coupled to a monopole
or instanton background, as well as multi-particle systems of Calogero 
type~\cite{calogero} (except for rational Calogero models~\cite{aacal}). 
Therefore, we have recently begun to develop this issue,
by analyzing the (one-dimensional) dihedral systems related to the three-particle
Calogero model~\cite{aa} as well as tow-dimensional oscillator-like systems
which are relevant to certain models of quantum rings and lenses~\cite{jctn}.

The key idea~\cite{hidden,aa} is to pick an integrable system with a
$2(N{-}1)$-dimensional {\sl compact\/} phase space,
\be
\cH=\cH(I_i), \qquad \{I_i,\Phi^0_j\}=\delta_{ij},\qquad
\Phi^0_i\in [0, 2\pi), \qquad
i,j=1,\ldots,N{-}1,
\ee
in terms of its action-angle variables, and add a {\sl radial\/} part to it,
\be
H=\frac{p^{2}_r}{2}+\frac{\cH(I_i)}{r^2}+V(r), \qquad \{p_r, r\}=1,\qquad
r\in[0,\infty) \quad\textrm{or}\quad [0, r_0) .
\label{2}\ee
Here, we introduced a radial coordinate~$r$ and momentum~$p_r$ and obtain
an extended model with $N$ degrees of freedom. The extended configuration space
is a cone over the original compact configuration space. If the latter is just
the sphere~$S^{N-1}$, we can obtain, in particular, the three model spaces of
constant curvature:
\bea
S^N &:& \quad r=r_0\sin\chi, \qquad\ \ p_r=r_0^{-1}p_\chi, \qquad
V(r)\to V(r_0\tan\chi), \label{3} \\[4pt]
\R^N &:& \quad r=r_0\chi, \qquad\qquad\, p_r=r_0^{-1}p_\chi, \qquad
V(r)\to V(r_0\chi), \label{5} \\[4pt]
H^N &:& \quad r=r_0\sinh\chi, \qquad p_r=r_0^{-1}p_\chi, \qquad
V(r)\to V(r_0\tanh\chi), \label{4}
\eea
where $r_0$ is the radial scale and $\{p_\chi,\chi\}=1$ is a dimensionless
canonical pair. Hence, for a particle on the sphere~$S^N$ (the sine-cone
over~$S^{N-1}$) or on the hyperboloid~$H^N$ (the hyperbolic cone over~$S^{N-1}$)
obe gets the Hamiltonians
\be
H=\frac{p_{\chi}^2}{2 r_0^2}+\frac{\cH}{r_0^2\sin^2\chi}+V(r_0\tan\chi)
\qquad\textrm{and}\qquad
H=\frac{p_{\chi}^2}{2 r_0^2}+\frac{\cH}{r_0^2\sinh^2\chi}+V(r_0\tanh\chi),
\label{6}\ee
respectively.

As an example, when $\cH$ defines the Landau problem, i.e.~a particle on $S^2$
moving in the magnetic field generated by a Dirac monopole located
at the center of sphere, we arrive at the particle on $\R^3$ under the influence
of this Dirac monopole.
The extended system remains integrable for two prominent choices of the radial
potential,
\be
V(r)=V_{\textrm{osc}}(r)=\sfrac12\omega^2 r^2 \qquad\textrm{and}\qquad
V(r)=V_{\textrm{cou}}(r)=-\frac{\gamma}{r},
\label{pot}\ee
with frequency $\omega$ and (positive) coupling~$\gamma$, respectively.
For~$\R^N$, these are the familiar oscillator and Coulomb potentials,
while for $S^N$ they have been named Higgs oscillator~\cite{higgs} and
Schr\"odinger-Coulomb~\cite{sch}, respectively.

If the system is spherically symmetric, i.e.~$S^{N-1}$ invariant,
the compact Hamiltonian~$\cH$ is just given by the SO($N$) Casimir function~$J^2$,
which defines the kinetic energy of a free particle on~$S^{N-1}$.
Deviations from spherical symmetry are encoded in~$\cH$. In other words,
replacing $J^2$ by the Hamiltonian of some compact $N{-}1)$-dimensional
integrable system defines a deformation of the $N$-dimensional oscillator
and Coulomb systems.

Particular examples with $N{=}2$ are the so-called
Tremblay-Turbiner-Winternitz (TTW)~\cite{TTW} and Post-Winternitz (PW)~\cite{PW}
models, defined on~$\R^2$, which have attracted some interest recently
(see, e.g.~\cite{TTWothers} and references therein). There the compact subsystem
on the circle~$S^1$ is just the famous P\"oschl-Teller system~\cite{flugge},
\be
\cH=\cH_{\textrm{PT}}=\frac{p^2_\varphi}{2}+\frac{k^2\alpha^2_1}{2\sin^2k\varphi}
+\frac{2k^2\alpha^2_1}{\cos^2k\varphi} \qquad\textrm{with}\qquad k\in\NN.
\label{PT}\ee

The paper is arranged as follows.
In Section~2 we give the necessary information on action-angle variables
and present the general formulae for the systems with Hamiltonians
(\ref{2}) and~(\ref{6}).
Explicit expressions for action-angle variables for these systems
with the potentials (\ref{pot}) are computed in Section~3.
In Section~4, we construct the spherical and pseudospherical generalizations
of the TTW and PW systems. We demonstrate the superintegrability of these systems
write down their hidden constants of motion.
In an Appendix we provide the action-angle variables for a free particle
on the $(N{-}1)$-dimensional sphere, which yields the complete set of
action-angle variables for the $N$-dimensional oscillator and Coulomb systems
as well as their spherical and pseudospherical analogs.

\setcounter{equation}{0}
\section{Action-angle variables}
The well-known Liouville theorem gives an exact criterion for the integrability
of an $N$-dimensional mechanical system: the existence of $N$ mutually commuting
constants of motion
\be
{\bf F}=(F_1{\equiv}H, F_2,\ldots,F_N) \qquad\textrm{with}\qquad
\{F_\mu,F_\nu\}=0 \qquad\textrm{for}\quad \mu,\nu=1,\ldots,N.
\ee
The theorem also states that the compact and connected level surfaces
$M_f=\big( (p_\mu,q_\nu): F_\mu =\textrm{const}\big)$ are diffeomorphic to
$N$-dimensional tori~$T^N$. The particle performs a free motion in the
natural angular coordinates ${\bf \Phi}=(\Phi_1,\ldots, \Phi_N)$
parameterizing such a torus. Their conjugate momenta, the so-called action
variables ${\bf I}=(I)\equiv(I_1,\ldots, I_N)$, are conserved and thus are
functions of the constants of motion, ${\bf I}={\bf I}({\bf F})$.
Hence, there exists a canonical transformation $({\bf p},{\bf q})\mapsto
({\bf I},{\bf \Phi})$, after which the Hamiltonian depends  on the (constant)
action variables only. Consequently, the equations of motion read
\be
\frac{d{\bf I}}{dt}=0 \qquad\textrm{and}\qquad
\frac{d{\bf \Phi}}{dt}=\frac{\partial H(I)}{\partial {\bf I}}
\qquad\textrm{for}\quad {\bf \Phi}\in T^N.
\ee

The general construction of action-angle variables proceeds as follows
\cite{arnold}.
First, fix a level surface ${\bf F}={\bf c}$.
Second, introduce the generating function~$S$
for the canonical transformation $({\bf p},{\bf q})\mapsto({\bf I},{\bf \Phi})$
as the indefinite integral
\be
S({\bf c},{\bf q})=\int_{{\bf F}={\bf c}} {\bf p}\cdot d{\bf q},
\label{sdef} \ee
where ${\bf p}={\bf p}({\bf c},{\bf q})$
by use of the equations of motion.
The action variables ${\bf I}$ can be obtained via
\be
I_\mu ({\bf c})=\frac{1}{2\pi}\oint_{\gamma_\mu} {\bf p}\cdot d{\bf q},
\ee
where $\gamma_\mu$ is some homology cycle of the level surface ${\bf F}={\bf c}$.
Third, invert these relations to get ${\bf c}={\bf c}({\bf I})$.
Fourth, the angle variables ${\bf \Phi}$ can be found from the expression
\be
{\bf \Phi}({\bf c},{\bf q})=
\frac{\partial S({\bf c}({\bf I}), {\bf q})}{\partial {\bf I}}
\Big|_{{\bf I}\to{\bf I}({\bf c})}.
\ee
Fifth, one may restore ${\bf c}={\bf c}({\bf p},{\bf q})$ to arrive at explicit
formulae for the canonical transformation.

Assume now that the compact subsystem is already formulated in terms of
action-angle variables $(I_i,\Phi_i^0)$, with $i=1,\ldots,N{-}1$, while the
radial part is given by~$(p_r,r)$.
We characterize the level sets by $(H{\equiv}E, I_i)$.
The generating function for the extended system~(\ref{2}) then reads
\be
S(E, I_i,r,\Phi_i^0)=\sqrt{2}\int\!dr\ \sqrt{E-\frac{\cH(I)}{r^2}-V(r)}\
+\ \sum_{i=1}^{N-1} I_i\Phi^0_i.
\label{11}\ee
>From this function we immediately get the action variables $I_i=I_i$ and
\be
I_r(E,I_i)=\frac{\sqrt{2}}{2\pi}\oint dr\ \sqrt{E-\frac{\cH(I)}{r^2}-V(r)}.
\label{12}\ee
The corresponding angle variables are given by
\be
\Phi_r=\frac{1}{\sqrt{2}}\frac{\partial E}{\partial I_r}
\int\!\frac{dr}{\sqrt{E-\frac{\cH(I)}{r^2}-V(r)}} \quad\textrm{and}\quad
\Phi_i=\Phi^0_i+
\frac{{\partial E}/{\partial I_i}}{\partial E/\partial I_r}\Phi_r-
\frac{1}{\sqrt{2}}\frac{\partial \cH(I)}{\partial I_i}
\int\!\frac{dr}{r^2\sqrt{E-\frac{\cH(I)}{r^2}-V(r)}}.
\label{13}\ee

Making  in (\ref{11}) and (\ref{13}) the replacements described
in (\ref{3}) or~(\ref{4}),
we shall get the system on the $N$-sphere or -pseudosphere.
Of course, for the full construction of the action-angle variables,
we need to provide the action-angle variables of the subsystem~$\cH$.

\setcounter{equation}{0}
\section{Deformed oscillator and Coulomb problems}
In this section we present the action-angle variables $(I_r, \Phi_r, \Phi_i)$
for the deformed oscillator and Coulomb systems given by the expressions
(\ref{2})--(\ref{pot}).
The action variables $I_i$ of the ``angular Hamiltonian'' $\cH$ remain unchanged,
while the angle variables $\Phi^0_i$ receive corrections, as seen in~(\ref{13}).
For notational simplicity we abbreviate $H({\bf p},{\bf q})=E$, put $r_0=1$
and drop the argument $I_i$ of~$\cH$. In the following, we list the results for
each of the six combinations in the table below:\\

\begin{tabular}{l|cc}
radial potential & oscillator & Coulomb \\ \hline
metric cone: \hfill $\R^N$ & Euclidean oscillator & Euclidean Coulomb \\
sine-cone: \hfill $S^N$ & spherical Higgs oscillator &
spherical Schr\"odinger-Coulomb \\
hyperbolic cone: $H^N$ & pseudospherical Higgs oscillator &
pseudospherical Schr\"odinger-Coulomb
\end{tabular}

\subsubsection*{Euclidean oscillator}
\be
H_{\textrm{osc}}=\frac{p_r^2}{2}+\frac{{\cal H}}{r^2} + \frac{\omega^2 r^2}{2}
=\omega \big(2 I_r + \sqrt{2 {\cal H}}\big),
\label{1osc} \ee
\be
I_r=\frac{E}{2\omega} -
\sqrt{\frac{{\cal H}}{2}} \qquad\textrm{and}\qquad
\Phi_r=-\arcsin\Big(\frac{E-r^2\omega^2}{\sqrt{E^2-2{\cal H}\omega ^2}}\Big),
\label{radosc}\ee
\be
\Phi_i =\Phi^0_i+
\frac{1}{2\sqrt{2\cH}}~\frac{\partial\cH}{\partial I_i}\biggl[\Phi_r-
\arcsin\Big(\frac{E\,r^2-2\cH}{r^2\sqrt{E^2-2\omega^2\cH}}\Big)\biggr].
\ee
\bigskip

\subsubsection*{Euclidean Coulomb}
\be
H_{\textrm{cou}}=\frac{p_r^2}{2}+\frac{{\cal H}}{r^2} - \frac{\gamma}{r}
= -\frac{\gamma^2}{2\big(I_r+\sqrt{2 {\cal H}}\big)^2},
\label{coulumb1} \ee
\be I_r = \frac{\gamma}{\sqrt{-2 E}}-\sqrt{2{\cal H}}
\qquad\textrm{and}\qquad
\Phi_r= -\frac2\gamma\sqrt{E {\cal H} - E\,r(E\,r+\gamma)}-
\arcsin\Big(\frac{2 E\,r+\gamma }{\sqrt{4 E {\cal H}+\gamma^2}}\Big),
\label{coulumb2}\ee
\be
\Phi_i=\Phi^0_i+\sqrt{\frac{2}{{\cal H}}}~\frac{\partial {\cal H}}{\partial I_i}
\biggl[\Phi_r-\frac{1}{2}\arcsin\Big(\frac{\gamma\,r-2\cH}{r\sqrt{4E\cH+\gamma^2}}
\Big)\biggr].
\ee
\bigskip

\subsubsection*{Spherical Higgs oscillator }
\be
H_{\textrm{s-higgs}}=\frac{p_{\chi}^2}{2}+\frac{{\cal H}}{\sin^2\chi} +
\frac{\omega^2\tan^2\chi}{2}
=\frac{1}{2}\left(2 I_\chi+\sqrt{2 {\cal H}}+\omega\right)^2-\frac{\omega^2}{2},
\label{1Higgs} \ee
\bea
&& I_\chi=\frac12 \Big(\sqrt{2 E + \omega^2}-\sqrt{2\cH}-\omega\Big)
\qquad\textrm{and} \nonumber\\[4pt]
&& \Phi_\chi = - 2 \arcsin \Big(
\frac{(2E+\omega^2)\cos 2\chi+2\cH-\omega^2}
{\sqrt{(2E+\omega^2)^2-2(2\cH+\omega^2)(2E+\omega^2)+(2\cH-\omega^2)^2}}\Big),
\eea
\be
\Phi_i=\Phi^0_i+
\frac{1}{2\sqrt{2 {\cal H}}}~\frac{\partial\cH}{\partial I_i}
\biggl[\Phi_\chi+ \arctan\Big(
\frac{(E+{\cal H}) \cos2\chi-E+3 {\cal H}}
{\sqrt{2\cH}~\sqrt{2E-4\cH-\omega^2-(4\cH-2\omega^2)\cos2\chi -
(2E+\omega^2)\cos^2 2\chi}}\Big) \biggr].
\ee
\bigskip

\subsubsection*{Spherical Schr\"{o}dinger-Coulomb}
\be
H_{\textrm{s-sch-cou}}=\frac{p_{\chi}^2}{2}+\frac{{\cal H}}{\sin^2\chi} -
\gamma\cot\chi
=\frac12\Big(I_\chi+\sqrt{2 {\cal H}}\Big)^2-
\frac{\gamma ^2}{2 (I_\chi+\sqrt{2 {\cal H}})^2},
\label{SCP} \ee
\be
I_\chi=\sqrt{E+\sqrt{E^2+\gamma^2}} -\sqrt{2 {\cal H}}
\qquad\textrm{and}\qquad
\Phi_\chi= \textrm{Im}\biggl[
\frac{2\sqrt{{\cal H}(E+\textrm{i}\gamma )}}
{\sqrt{E+\sqrt{E^2+\gamma^2}}}\log\zeta \biggr]
\ee
\be
\Phi_i=\Phi^0_i+
\frac{1}{\sqrt{2 {\cal H}}}~\frac{\partial {\cal H}}{\partial I_i}
\biggl[\Phi_\chi+ \arcsin\Big(\frac{2 {\cal H} \cot\chi-\gamma}
{\sqrt{4 (E-{\cal H}) {\cal H}+\gamma ^2}}\Big)\biggr],
\ee
\be
\textrm{where}\qquad
\zeta = \frac{4}{\sqrt{{\cal H}}} \textrm{e}^{\textrm{i}(\chi+\frac{\pi}{2})}
\sin\chi\Big(1+\sqrt{
\vphantom{\Big|}E-\smash{\frac{{\cal H}}{\sin^2\chi}}+\gamma\cot\chi}\Big)+
\frac{(4{\cal H}-2 \textrm{i}\gamma)\sqrt{E+\textrm{i}\gamma}}
{\sqrt{{\cal H}(E^2+\gamma^2)}}.
\ee
\bigskip

\subsubsection*{Pseudospherical Higgs oscillator}
\be
H_{\textrm{ps-higgs}}=\frac{p_{\chi}^2}{2}+\frac{{\cal H}}{\sinh^2\chi} +
\frac{\omega^2\tanh^2\chi}{2}
=\frac{\omega^2}{2}-\frac{1}{2}\big(2I_\chi+\sqrt{2 {\cal H}}-\omega\big)^2,
\label{2Higgs} \ee
\bea
&& I_\chi=\frac{1}{2}\Big(\omega-\sqrt{2 {\cal H}} -\sqrt{\omega^2 -2 E}\Big),
\qquad\textrm{and} \nonumber \\[4pt]
&& \Phi_\chi = - 2 \arctan\Big(\frac{(1-\sqrt{1-\eta^2})
\left(E+{\cal H}-\omega^2\right)}{\eta ~ \omega  \sqrt{\omega^2 - 2
E}}+\frac{\sqrt{(E+{\cal H})^2-2 {\cal H} \omega^2}}{\omega
\sqrt{\omega^2 - 2 E}}\Big),
\eea
\be
\Phi_i=\Phi^0_i+\frac{1}{2\sqrt{2{\cal H}}}~\frac{\partial{\cal H}}{\partial I_i}
\biggl[\Phi_\chi-2 \arctan\Big( \frac{(1-\sqrt{1-\eta^2})
\left(E+{\cal H}\right)}{\eta~\omega\sqrt{2{\cal H}}}+
\frac{\sqrt{(E+{\cal H})^2-2 {\cal H} \omega^2}}{\omega \sqrt{2 {\cal H}}}
\Big)\biggr],
\ee
\be \textrm{where}\qquad\qquad
\eta=\frac{\omega^2\tanh^2\chi-(E+\cH)}{\sqrt{(E+\cH)^2-2\cH\omega^2}}.
\ee
\bigskip

\subsubsection*{Pseudospherical Schr\"{o}dinger-Coulomb}
\be
H_{\textrm{ps-sch-cou}}=\frac{p_{\chi}^2}{2}+\frac{{\cal H}}{\sinh^2\chi} -
\gamma\coth\chi
=-\frac12\Big(I_\chi+\sqrt{2 {\cal H}}\Big)^2-
\frac{\gamma ^2}{2(I_\chi+\sqrt{2 {\cal H}})^2},
\label{PsSCP} \ee
\bea
&& I_\chi=\frac1{\sqrt{2}}\Big(\sqrt{-E+\gamma}-\sqrt{-E-\gamma}-2\sqrt{\cH}\Big)
\qquad\textrm{and} \nonumber \\[4pt]
&& \Phi_\chi=\frac{\sqrt{-E+\gamma}}{\sqrt{2}(\sqrt{-E-\gamma}-\sqrt{-E+\gamma })}
\arctan\Big(\frac{\sqrt{4{\cal H}(E+{\cal H})+\gamma^2}+(\gamma-2{\cal H})\eta}
{2 \sqrt{{\cal H} (-E-\gamma ) }\sqrt{1-\eta ^2}}\Big) \nonumber \\[4pt]
&&\qquad- \frac{\sqrt{-E-\gamma}}{\sqrt{2}(\sqrt{-E-\gamma}-\sqrt{-E+\gamma })}
\arctan\Big(\frac{\sqrt{4{\cal H}(E+{\cal H})+\gamma^2}+(\gamma+2{\cal H})\eta}
{2 \sqrt{{\cal H} (-E+\gamma ) }\sqrt{1-\eta ^2}}\Big),
\eea
\be
\Phi_i=\Phi^0_i+
\frac{1}{\sqrt{2 {\cal H}}}~\frac{\partial {\cal H}}{\partial I_i}
\biggl[\Phi_\chi+ \frac{1}{\sqrt{2}}\arcsin\Big(
\frac{2\cH\cot\chi-\gamma}{\sqrt{4(E+\cH)\cH+\gamma^2}}\Big)\biggr].
\ee
\bigskip

\setcounter{equation}{0}
\section{Generalizations of the Tremblay-Turbiner-Winternitz system}
As mentioned in the Introduction, action-angle variables elegantly
explain the superintegrability of the recently suggested deformation of the
two-dimensional oscillator system introduced by
Tremblay-Turbiner-Winternitz (TTW)~\cite{TTW} and also of the
Coulomb versions treated by Post-Winternitz (PW)~\cite{PW}.
They also allow us to construct
analogous deformations of other superintegrable systems.

Our generalizations of the TTW and PW systems are defined
by (\ref{2})--(\ref{pot}) with $N{=}2$,
where the one-dimensional ``angular" Hamiltonian $\cH$ is
given by the generalized P\"oschl-Teller system on the circle
(\ref{PT})~\cite{flugge}. The action-angle variables of this subsystem
are given by \cite{aa}
\be
I_{\textrm{PT}}=\sfrac{1}{k}\sqrt{2\cH_{\textrm{PT}}}-(\alpha _1+\alpha _2)
\qquad\textrm{and}\qquad
\Phi_{\textrm{PT}}=\sfrac12\arcsin\Bigl\{\sfrac{1}{a}\bigl[\cos{2k\varphi}
+b\bigr]\Bigr\},
\label{che}\ee
where
\be
a=\sqrt{1-\frac{{k}^2(\alpha_1^2+\alpha_2^2)}{\cH_{\textrm{PT}}}
+\left(\frac{{k}^2(\alpha _1^2-\alpha _2^2)}{2\cH_{\textrm{PT}}}\right)^2}
\qquad\textrm{and}\qquad
b=\frac{{k}^2(\alpha_2^2-\alpha_1^2)}{2\cH_{\textrm{PT}}},
\label{definitions} \ee
so that the Hamiltonian reads
\be
\cH_{\textrm{PT}}=\sfrac12 (k{\tilde I_{\textrm{PT}}})^2
\qquad\textrm{with}\qquad
{\tilde I}_{\textrm{PT}}\equiv I_{PT}+\alpha_1+\alpha_2 \
\in[\alpha_1{+}\alpha_2,\infty).
\ee
Clearly, in action-angle variables, the P\"oschl-Teller Hamiltonian coincides
with the Hamiltonian of a free particle on a circle of radius~$k$,
but with a different domain for the action variable. Hence, choosing the
potential in~(\ref{2}) to be of oscillator or Coulomb type, the extended system
will be superintegrable. More precisely, in the variables $\big(p_r,r,
I_{\textrm{PT}}(p_\varphi,\varphi),\Phi_{\textrm{PT}}(p_\varphi,\varphi)\big)$,
this system takes the form of a conventional two-dimensional
oscillator or Coulomb system on the cone. Hence, for rational
values of~$k$ these systems possess hidden symmetries.
For the oscillator case, the hidden constants of motion
have been constructed in~\cite{gonera}. Here, we extend their results
to the Coulomb case~\cite{PW} as well as to the TTW- and PW-like systems
on spheres and pseudospheres. 

For the three spaces of constant curvature and for the oscillator
potential, the action-angle Hamiltonians are
\be
H_\omega=
\left\{\begin{array}{ccc}
\omega\,(2 I_\chi + k{\tilde I}_{\textrm{PT}}) \qquad\!\qquad
& \qquad {\rm for}& \R^2\\[2pt] \phantom{-}
\sfrac12(2I_\chi+k{\tilde I}_{\textrm{PT}}+\omega)^2-\sfrac{\omega^2}{2}
& \qquad {\rm for}& S^2\\[2pt]
-\sfrac12(2I_\chi+k{\tilde I}_{\textrm{PT}}-\omega)^2+\sfrac{\omega^2}{2}
& \qquad {\rm for}& H^2
\end{array} \right.
\ee
and depend only on the combination $2 I_\chi{+}k{\tilde I}_{\textrm{PT}}$.
Thus, the evolution of the angle variables is given by
\be
\Phi_\chi(t) = 2\,\Omega\,t \qquad\textrm{and}\qquad
\Phi_\varphi(t) = k\,\Omega\,t \qquad\textrm{with}\qquad
\Omega=\frac{d H_\omega}{d(2 I_\chi{+}k{\tilde I}_{\textrm{PT}})}.
\ee
For rational values of~$k$ the trajectories are closed.
It then follows that the hidden constant of motion is
\be
I_{\textrm{hidden}}=\cos \big(m\Phi_\chi{-}2n\Phi_\varphi\big)
\qquad\textrm{for}\quad k=m/n.
\ee
Explicitly, this hidden constant of motion reads:
\subsubsection*{Euclidean TTW system}

$$
I_{add}=CM_m\left(\frac{E r^2-2{\cal H}_{PT}}{r^2\sqrt{E^2-2  \omega^2{\cal H}_{PT}}}\right) CM_n \left(\sfrac{1}{a}\left[\cos{2k\varphi}
+b\right]\right) 
$$
\be
\phantom{.}\qquad
+SM_m\left(\frac{E r^2-2{\cal H}_{PT}}{r^2\sqrt{E^2-2  \omega^2{\cal H}_{PT}}}\right) SM_n \left(\sfrac{1}{a}\left[\cos{2k\varphi}
+b\right]\right)
\ee
where we denoted
$$
CM_n(x)=\cos(n \arcsin x)=\sum_{i=0}^{[\frac n2]}(-1)^i C_n^{2 i} x^{2 i}\sqrt{1-x^2}^{n-2 i},
$$
\be
SM_n(x)=\sin(n \arcsin x)=\sum_{i=0}^{[\frac{n-1}{2}]}(-1)^i C_n^{2 i+1} x^{2 i+1}\sqrt{1-x^2}^{n-2 i-1}.
\ee

\subsubsection*{Spherical  TTW system }
\be
I_{add}=CM_m\left(\frac{\xi}{\sqrt{\xi^2+1}}\right) CM_n \left(\sfrac{1}{a}\left[\cos{2k\varphi}+b\right]\right) -
 SM_m\left(\frac{\xi}{\sqrt{\xi^2+1}}\right) SM_n \left(\sfrac{1}{a}\left[\cos{2k\varphi}+b\right]\right)
\ee
where
\be
\xi=\frac{(E+{\cal H}_{PT}) \cos2\chi-E+3 {\cal H}_{PT}}
{\sqrt{2 {\cal H}_{PT}} \sqrt{2 E-4 {\cal H}_{PT}-\omega^2- \left(4 {\cal H}_{PT}-
2 \omega ^2\right)\cos2\chi - \left(2 E+\omega ^2\right) \cos^2 2\chi}}.
\ee

\subsubsection*{Pseudospherical TTW system}

\be
I_{add}=CM_{2m}\left(\frac{\xi}{\sqrt{\xi^2+1}}\right) CM_n \left(\sfrac{1}{a}\left[\cos{2k\varphi}+b\right]\right) +
 SM_{2m}\left(\frac{\xi}{\sqrt{\xi^2+1}}\right) SM_n \left(\sfrac{1}{a}\left[\cos{2k\varphi}+b\right]\right)
\ee

where
$$
\xi=\frac{\sqrt{(E+{\cal H}_{PT})^2-2 {\cal H}_{PT} \omega^2}-\sqrt{2(E+{\cal H}_{PT})\omega^2
\tanh^2\chi-\omega^4
\tanh^4\chi-2 {\cal H}_{PT} \omega^2}}{\omega^2
\tanh^2\chi-(E+{\cal H}_{PT})} \frac{
E+{\cal H}_{PT}}{ \omega  \sqrt{2 {\cal H}_{PT}}}
$$
\be
+\frac{\sqrt{(E+{\cal H}_{PT})^2-2 {\cal H}_{PT} \omega^2}}{\omega
\sqrt{2 {\cal H}_{PT}}}. 
\qquad\qquad\qquad\qquad\qquad\qquad\qquad\qquad\qquad\qquad\qquad\phantom{.}
\ee
Thus, choosing the Higgs oscillator on the (pseudo)sphere, we
get a superintegrable (pseudo)spherical analog of the TTW~oscillator.\\

The construction of superintegrable deformations of the Coulomb system,
i.e.~the PW~model and its generalization to the (pseudo)spherical environment,
proceeds completely similarly. The Hamiltonians
\be
H_\gamma=
\left\{\begin{array}{ccc}
-\sfrac{\gamma^2}{2} (I_\chi + k{\tilde I}_{\textrm{PT}})^{-2}
\qquad\qquad\qquad\qquad
& \qquad {\rm for }& \R^2\\[2pt]
-\sfrac{\gamma^2}{2} (I_\chi + k{\tilde I}_{\textrm{PT}})^{-2}
+\sfrac12(I_\chi+k{\tilde I}_{\textrm{PT}})^2
& \qquad {\rm for}& S^2\\[2pt]
-\sfrac{\gamma^2}{2} (I_\chi + k{\tilde I}_{\textrm{PT}})^{-2}
-\sfrac12(I_\chi+k{\tilde I}_{\textrm{PT}})^2
& \qquad {\rm for}&  H^2
\end{array} \right.
\ee
depend only on the combination $I_\chi{+}k{\tilde I}_{\textrm{PT}}$,
and for rational $k=m/n$ the trajectories are closed, supporting
\be
I_{\textrm{hidden}}=\cos \big(m\Phi_\chi{-}n\Phi_\varphi\big).
\ee
Explicitly this constant of motion reads:

\subsubsection*{Euclidean PW system}

$$
I_{add}=CM_{2 m}\left(\frac{\gamma r -2 {\cal H}_{PT} }{r \sqrt{4 E {\cal H}_{PT}+\gamma ^2}}\right) CM_n\left(\sfrac{1}{a}\bigl[\cos{2k\varphi}+b\bigr]\right)
$$
\be
\phantom{.}\qquad
+SM_{2 m}\left(\frac{\gamma r -2 {\cal H}_{PT} }{r \sqrt{4 E {\cal H}_{PT}+\gamma ^2}}\right) SM_n\left(\sfrac{1}{a}\bigl[\cos{2k\varphi}+b\bigr]\right).
\ee

\subsubsection*{Spherical PW system}
$$
I_{add}=CM_{2 m}\left(\frac{2 {\cal H}_{PT} \cot\chi-\gamma}{\sqrt{4 (E-{\cal H}_{PT}) {\cal H}_{PT}+\gamma ^2}}\right) CM_n\left(\sfrac{1}{a}\bigl[\cos{2k\varphi}
+b\bigr]\right)
$$
\be
\phantom{.}\qquad
-SM_{2 m}\left(\frac{2 {\cal H}_{PT} \cot\chi-\gamma}{\sqrt{4 (E-{\cal H}_{PT}) {\cal H}_{PT}+\gamma ^2}}\right) SM_n\left(\sfrac{1}{a}\bigl[\cos{2k\varphi}
+b\bigr]\right).
\ee

\subsubsection*{Pseudospherical PW system}

$$
I_{add}=CM_{2 m}\left(\frac{2 {\cal H}_{PT} \coth\chi-\gamma}{\sqrt{4 (E+{\cal H}_{PT}) {\cal H}_{PT}+\gamma ^2}}\right) CM_n\left(\sfrac{1}{a}\bigl[\cos{2k\varphi}
+b\bigr]\right)
$$
\be
\phantom{.}\qquad
-SM_{2 m}\left(\frac{2 {\cal H}_{PT} \coth\chi-\gamma}{\sqrt{4 (E+{\cal H}_{PT}) {\cal H}_{PT}+\gamma ^2}}\right) SM_n\left(\sfrac{1}{a}\bigl[\cos{2k\varphi}
+b\bigr]\right).
\ee
Thus, choosing the Schr\"odinger-Coulomb system on the (pseudo)sphere,
we get a superintegrable (pseudo)spherical analog of the PW~model.

\section{Summary and outlook}
We have presented the action-angle variables for a particular class of
integrable deformations of $N$-dimensional oscillator and Coulomb systems,
on flat space as well as on the sphere and the pseudosphere.
These integrable systems were obtained by replacing the
angular part of the ordinary (spherically symmetric) oscillator or Coulomb
systems with a suitable compact integrable system formulated in terms of its
action-angle variables.

As the application with $N{=}2$, we constructed the spherical and pseudospherical
generalization of the Tremblay-Turbiner-Winternitz (TTW) and Post-Winternitz (PW)
models, demonstrated their superintegrability and computed their hidden constant
of motion. For completeness, we also provided the action-angle variables of the
undeformed angular subsystem, i.e.~for a free particle moving on the
$(N{-}1)$-sphere.

An obvious task is to extend the above example to higher dimensions,
by employing generalizations of the P\"oschl-Teller systems to construct
higher-dimensional analogs of the TTW and PW models.

\bigskip

\noindent
{\large\bf Acknowledgments.} \quad
This work was partially supported by
Volkswagen Foundation grant I/84 496
and by the grants SCS 11-1c258 and SCS-BFBR 11AB-001 of the
Armenian State Committee of Science.

\setcounter{equation}{0}
\def\theequation{A.\arabic{equation}}
\section*{Appendix: Free particle on $S^{N-1}$}
In this Appendix we recollect the  action-angle variables for
the ``angular Hamiltonian" ${\cal H}_{PT}$ appearing in every spherically symmetric
$N$-dimensional system and defining the free motion of a particle on $S^{N-1}$
with radius $r_0=1$.
It is given by the Casimir function  $L^2_N$ of SO($N$),
\be
\cH=\sfrac12 L^2_N.
\label{l2}\ee

The embedding of the unit $(N{-}1)$-sphere into $\R^N$ is given
by a set of polar coordinates,
\be
\left.
\begin{aligned}
x_1&=s_{N-1}\,s_{N-2}\cdots s_3\,s_2\,s_1 \quad \\
x_2&=s_{N-1}\,s_{N-2}\cdots s_3\,s_2\,c_1 \\
x_3&=s_{N-1}\,s_{N-2}\cdots s_3\,c_2 \\[-4pt]
&\vdots \\[-4pt]
x_{N-1}&=s_{N-1}\,c_{N-2} \\
x_N&=c_{N-1}
\end{aligned}
\right\} \qquad\qquad
\begin{aligned}
\textrm{with}\qquad
& s_k:=\sin\theta_k \quad\textrm{and}\quad c_k:=\cos\theta_k \\[4pt]
\textrm{for}\qquad
& \theta_1\in[0,2\pi)\ ,\quad \theta_{k>1}\in[0,\pi) \\[4pt]
\textrm{and}\qquad
& k=1,2,\ldots,N{-}1.
\end{aligned}
\ee
In these coordinates, we have the recursion
\be
L^2_N = p_{N-1}^2+\frac{L^2_{N-1}}{s^2_{N-1}}
\label{reccur}\ee
where
$p_{N-1}$ is the momentum conjugate to $\theta_{N-1}$.
It is easy to see that the $L^2_k$ for $k=1,\ldots,N$ are
in involution with each other and, therefore, can be used for constructing
action-angle variables.
Each variable $\theta_k$ defines an independent homology cycle~$S^1_k$
of the torus~$T^N$.
The level surfaces $L^2_k=\textrm{constant}=:j_k$ are diffeomorphic to~$T^N$.

Following the standard procedure we should compute the $N$ integrals
\be
I_k=\frac{1}{2\pi}\oint\limits_{S^1_k}{\bf p}\cdot d{\bf q}
=\frac{1}{2\pi}\oint\limits_{S^1_k}p_k\,d\theta_k=
\int\limits^{\theta_k^+}\limits_{\theta_k^-}
\sqrt{j_k-\frac{j_{k-1}}{\sin^2{\theta_k}}}\;d\theta_k,
\label{gends}
\ee
where in the second equality we used that the $\theta_k$ are mutually orthogonal
and the cycles $S_k^1$ are independent.
The integration ranges $[\theta_k^-,\theta_k^+]$ are determined from
the condition that the radicants should be non-negative.
Substituting
\be
u_k=\sqrt{\frac{j_k}{j_k-j_{k-1}}}\,\cos{\theta_k},
\ee
we arrive at
\be
I_k=2\,\frac{j_k{-}j_{k-1}}{2\pi\sqrt{j_k}}\int\limits_{-1}^1
\frac{\sqrt{1-u_k^2}\;du_k}{1-\frac{j_k-j_{k-1}}{j_k}u_k^2}=
\sqrt{j_k}-\sqrt{j_{k-1}},
\qquad\textrm{so that}\qquad
\sqrt{j_k}=\sum\limits_{m=1}^k I_m.
\label{acv}\ee
For the generating function we obtain
\be
S=\sum\limits_{l=1}^{N-1}S_l\qquad\textrm{where}\qquad
S_l=\int\!d\theta_l\ {\sqrt{\biggl(\sum\limits_{m=1}^l I_m\biggr)^2-\
\sin^{-2}\theta_l \biggl(\sum\limits_{m=1}^{l-1}I_m\biggr)^2}},
\ee
from which we get the angle variables
\be
\Phi^0_k=\frac{\partial S}{\partial I_k}=
\frac{\partial S_k}{\partial I_k}+
\sum\limits_{l=k+1}^{N-1}\frac{\partial S_l}{\partial I_k}=
\int\!\frac{\sqrt{j_k}\;d\theta_k}{\sqrt{j_k-\frac{j_{k-1}}{\sin^2{\theta_k}}}}+
\sum\limits_{l=k+1}^{N-1}
\int\!\frac{d\theta_l}{\sqrt{j_l-\frac{j_{l-1}}{\sin^2{\theta_l}}}}
\biggl(\sqrt{j_l}-\frac{\sqrt{j_{l-1}}}{\sin^2{\theta_l}}\biggr).
\label{a14} \ee
The first integral can be included in the first part of the sum (as $l{=}k$),
which yields
\be
\sum\limits_{l=k}^{N-1}\int\!\frac{\sqrt{j_l}\;d\theta_l}
{\sqrt{j_l-\frac{j_{l-1}}{\sin^2{\theta_l}}}}=
\sum\limits_{l=k}^{N-1}\arcsin{u_l}.
\ee
After the substitution and abbreviation
\be
u_l=\frac{2t_l}{(1{+}t_l)^2} \qquad\textrm{and}\qquad
a=\sqrt{\frac{j_l-j_{l-1}}{j_l}}\ <1,
\ee
respectively, the second part of the sum in (\ref{a14}) becomes
\be
\sum\limits_{l=k+1}^{N-1}\!\!\sqrt{\frac{j_{l-1}}{j_l}}
\int\!\!dt_l\biggl[\frac{1}{(t_l{-}a)^2+1-a^2}+\frac{1}{(t_l{+}a)^2+1-a^2}\biggr]
=\sum\limits_{l=k+1}^{N-1}\!\sqrt{\frac{j_{l-1}}{j_l(1{-}a^2)}}
\left[\arctan{\frac{t_l{-}a}{\sqrt{1{-}a^2}}}
+\arctan{\frac{t_l{+}a}{\sqrt{1{-}a^2}}}\right].
\ee
Pulling all together, we finally find
\be
\Phi^0_k=\sum\limits_{l=k}^{N-1}\arcsin{u_l}+\sum\limits_{l=k+1}^{N-1}
\arctan\biggl({\sqrt{\frac{j_{l-1}}{j_l}}\frac{u_l}{\sqrt{1-u_l^2}}}\biggr).
\label{anv}\ee

To summarize, the action-angle variables for a free particle on $S^{N-1}$
are given by (\ref{acv}) and~(\ref{anv}), with
$j_l=L^2_l(p_1,\ldots,p_l,\theta_1,\ldots,\theta_l)$.
The angular Hamiltonian (\ref{l2}) can be expressed as
\be
\cH=\sfrac12 L^2_N=\frac12\biggl(\sum\limits_{m=1}^{N-1} I_m\biggr)^2.
\label{free}\ee

\end{document}